\documentclass[12pt]{iopart}

\usepackage{iopams}
\usepackage{amssymb}
\usepackage{graphicx}
\usepackage{bm}

\begin{document}

\title{Orbital-based Scenario for Magnetic Structure of
Neptunium Compounds}

\author{Hiroaki Onishi and Takashi Hotta}

\address{Advanced Science Research Center,
Japan Atomic Energy Research Institute,
Tokai, Ibaraki 319-1195, Japan}

\begin{abstract}
In order to understand a crucial role of orbital degree of freedom
in the magnetic structure of recently synthesized neptunium
compounds NpTGa$_5$ (T=Fe, Co, and Ni), we propose to discuss
the magnetic phase of an effective two-orbital model,
which has been constructed based on a $j$-$j$ coupling scheme
to explain the magnetic structure of uranium compounds UTGa$_5$.
By analyzing the model with the use of numerical technique
such as exact diagonalization,
we obtain the phase diagram including several kinds of magnetic states.
An orbital-based scenario is discussed to understand the change
in the magnetic structure among C-, A-, and G-type antiferromagnetic
phases, experimentally observed in NpFeGa$_5$, NpCoGa$_5$, and NpNiGa$_5$.
\end{abstract}


\maketitle

%
%
\section{Introduction}

Recently it has been widely recognized that orbital degree of freedom
plays an essential role for the understanding of novel magnetism
in transition metal oxides \cite{Dagotto}.
A typical material is manganese oxide,
exhibiting remarkable colossal magneto-resistance phenomena \cite{Tokura}.
Due to competition and interplay among spin, charge, and orbital
degrees of freedom, rich phase diagram has been revealed \cite{Dagotto},
but a recent trend is to unveil further new phases
both from experimental and theoretical investigations.
In fact, even in undoped RMnO$_3$ (R=rare earth lanthanide ions),
a novel antiferromagnetic (AF) phase called the ``E-type'' spin structure
has been reported as the ground state for R=Ho \cite{Munoz,Kimura}.
Here we follow the definitions of spin structure in Ref.~\cite{Wollan}.
The origin of the E-AF phase has been clarified theoretically
\cite{Hotta-E} based on a band-insulator scenario in the
$e_{\rm g}$-orbital systems \cite{Hotta-IJMPB,Hotta-Topology,Hotta-CE}.
It should be noted that the ground state of undoped manganites was just
considered to be well understood, since for most R-ions, the A-type AF
insulating phase appears with the C-type ordering
of the $(3x^2$$-$$r^2)$- and $(3y^2$$-$$r^2)$-orbitals
\cite{Mizokawa,Solovyev,Koshibae,Shiina,Ishihara,Sawada,Feinberg,
Maezono,Feiner,Horsch,Brink,Benedetti,Hotta-A,Allen,Capone}.
Moreover, also for the half-doped manganite ${\rm La_{0.5}Ca_{0.5}MnO_3}$,
a charge-ordered ferromagnetic (FM) phase has been found in experiments
\cite{Loudon,Mathur},
as predicted theoretically \cite{Yunoki,Hotta-Stripe}.
These facts clearly indicate the importance of both experimental
and theoretical efforts to unveil new phases in manganites
in addition to the explanation of the complex phases already observed.
Such efforts have also been made to find new phases in
other transition metal oxides, for instance,
ruthenates \cite{Hotta-Ru,Nakamura} and nickelates \cite{Hotta-Ni}.

A trend to seek for new magnetic as well as superconducting phases
has been also found in the $f$-electron system, which is another type
of spin-charge-orbital complex.
Among many kinds of $f$-electron materials, in recent years,
$f$-electron compounds with HoCoGa$_5$-type tetragonal
crystal structure [see Fig.~1(a)], frequently referred to as ``115'',
have been intensively investigated both in experimental and theoretical
research fields of condensed matter physics.
Such vigorous activities are certainly motivated by
high superconducting transition temperature $T_{\rm c}$
observed in some 115 compounds.
Especially, amazingly high value of $T_{\rm c}$=18.5K has been reported
in PuCoGa$_5$ \cite{PuCoGa5,Sarrao,Wastin} and the mechanism has been
discussed theoretically \cite{Hotta-jj,Maehira}.
Among 115 compounds, interesting magnetic properties have been
reported for UTGa$_5$, where T is a transition metal ion
\cite{U115-1a,U115-1b,U115-1c,U115-1d,U115-2,U115-3,U115-4,U115-5,
U115-6,U115-7,U115-8,U115-9}.
In particular, neutron scattering experiments have revealed that
UNiGa$_5$ exhibits the G-type AF phase,
while UPdGa$_5$ and UPtGa$_5$ have the A-type AF state
\cite{U115-5,U115-9}.
Note that G-type indicates a three-dimensional N\'eel state,
while A-type denotes a layered AF structure in which
spins align ferromagnetically in the $ab$ plane and
AF along the $c$ axis \cite{Wollan}.
It is quite interesting that the magnetic structure is different
for U-115 compounds which differ only by the substitution
of transition metal ions.

\begin{figure}[t]
\begin{center}
\includegraphics[width=0.9\linewidth]{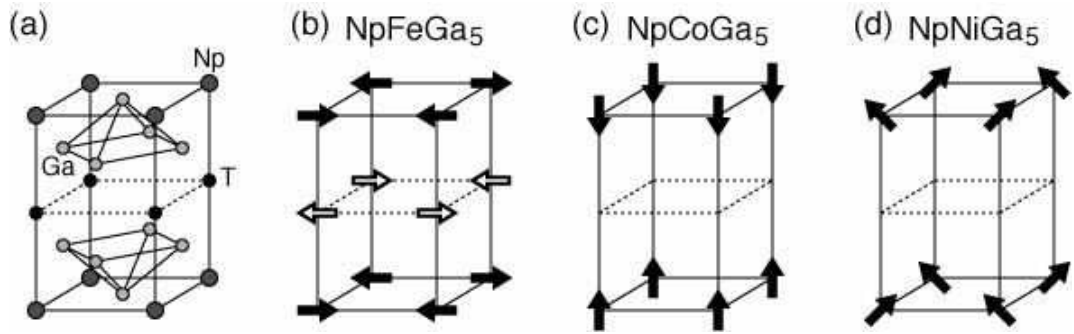}
\caption{%
(a) Crystal structure of NpTGa$_5$.
Schematic views of magnetic structures composed of magnetic moments
of Np ions for (b) NpFeGa$_5$, (c) NpCoGa$_5$, and (d) NpNiGa$_5$.
For NpFeGa$_5$, magnetic moments at Fe sites are also depicted.}
\end{center}
\end{figure}

Quite recently, 115 compounds including neptunium have been
skillfully synthesized and several kinds of physical quantities
have been successfully measured
\cite{Colineau,Aoki-Ni,Aoki-Co1,Aoki-Co2,Honda,Yamamoto,Homma,Metoki}.
Especially, the de Haas-van Alphen (dHvA) effect has been observed
in NpNiGa$_5$ \cite{Aoki-Ni}, which is the first observation of
dHvA signal in transuranium compounds.
For NpCoGa$_5$, the dHvA oscillations have been also detected
and a couple of cylindrical Fermi surfaces are found \cite{Aoki-Co2}.
For NpFeGa$_5$, the magnetic moment at Fe site has been suggested
in neutron scattering experiments \cite{Metoki} and it has been
also detected by $^{57}$Fe M\"ossbauer spectroscopy \cite{Homma}.
Interestingly enough, the magnetic structure of Np-115 compounds
also depends sensitively on transition metal ion \cite{Honda,Metoki}:
C-AF for NpFeGa$_5$, A-AF for NpCoGa$_5$, and G-AF for NpNiGa$_5$,
as shown in Figs.~1(b)-(d).
Note that for NpNiGa$_5$, the G-AF structure is composed of canted
Np moments and the peak in the neutron scattering intensity grows
$after$ the FM transition occurs \cite{Honda,Metoki}.
In any case, it is characteristic of U-115 and Np-115 compounds
that the magnetic properties are sensitive to the choice of
transition metal ions.

The appearance of several kinds of AF states reminds us
of the magnetic phase diagram of manganites and thus,
we envisage a scenario to understand the complex magnetic structure
of actinide compounds based on an orbital degenerate model
similar to that of manganites.
However, one must pay close attention to the meanings of
``spin'' and ``orbital'' in $f$-electron systems.
Since they are tightly coupled with each other through
a strong spin-orbit interaction, distinguishing them is not
straightforward in comparison with $d$-electron systems.
This point can create serious problems when we attempt to
understand microscopic aspects of magnetism and
superconductivity in $f$-electron compounds.
Thus, it is necessary to carefully define the terms ``orbital''
and ``spin'' for $f$ electrons in a microscopic discussion of
magnetism and superconductivity in actinide compounds.

In order to overcome such problems, we have proposed to employ
a $j$-$j$ coupling scheme to discuss $f$-electron systems
\cite{Hotta-jj}.
Here we stress the advantages of the $j$-$j$ coupling scheme.
First, it is quite convenient for the inclusion of many-body effects
using standard quantum-field theoretical techniques,
since individual $f$-electron states are clearly defined.
In contrast, in the $LS$ coupling scheme we cannot use such standard
techniques, since Wick's theorem does not hold.
Second we can, in principle, include the effects of valence fluctuations.
In some uranium compounds, the valence of the uranium ion is neither
definitely U$^{3+}$ nor U$^{4+}$, indicating that the $f$-electron
number takes a value between 2 and 3.
In the $j$-$j$ coupling scheme this is simply regarded
as the average number of $f$ electron per uranium ion.

In this paper, we attempt to explain the complex magnetic structure
of Np-115 based on the effective two-orbital model, which has been
constructed from the $j$-$j$ coupling scheme in order to understand
the magnetic properties of U-115 compounds \cite{Hotta-U115}.
We depict the Fermi surfaces of the kinetic term of
$5f$ electron based on the $j$-$j$ coupling scheme
in comparison with those obtained from the dHvA experiments.
The agreement seems to be fairly well, but the present itinerant picture
should be critically discussed.
Then, we apply the exact diagonalization technique to the model
in a small 2$\times$2$\times$2 cube to obtain a clue to understand
the complex magnetic structure of NpTGa$_5$.
The phase diagrams are found to include G-, A-, and C-type AF states,
consistent with experimental observations in Np-115.

The organization of this paper is as follows.
In Sec.~2, we will introduce an effective Hamiltonian,
called the $\Gamma_8$ model, for actinide 115 systems.
In Sec.~3, we show the calculated results of the model Hamiltonian.
The Fermi-surface structure is discussed in comparison
with dHvA experimental results.
We show numerical results for the magnetic structure obtained
by using exact diagonalization technique.
Finally, in Sec.~4, future developments are discussed and
the paper is summarized.
Throughout the paper, we use units such that $\hbar$=$k_{\rm B}$=1.

%
%
\section{Model Hamiltonian}

In this section, we show an effective model based on the $j$-$j$
coupling scheme for Np-115 compounds, which is the same as the model
for U-115 \cite{Hotta-U115}.
We emphasize that the model with orbital degree of freedom is
applicable to actinide 115 materials in common.

\subsection{Local $f$-electron state}

In order to construct a microscopic Hamiltonian for $f$-electron systems,
it is necessary to include simultaneously the itinerant nature of
$f$ electrons as well as strong electron correlation and
the effect of crystalline electric field (CEF).
For the purpose, we have proposed to use the so-called $j$-$j$
coupling scheme \cite{Hotta-jj}.
As shown in Fig.~2, we include the spin-orbit coupling so as
to define the state labeled by the total angular momentum.
For $f$ orbitals with $\ell$=3, we immediately obtain an octet with
$j$=7/2(=3+1/2) and a sextet with $j$=5/2(=3$-$1/2),
which are well separated by the spin-orbit interaction.
In general, the spin-orbital coupling is as large as 1eV
in actinide elements and thus, in the $j$-$j$ coupling scheme,
we consider only the $j$=5/2 sextet.

In actual compounds, due to the electrostatic potential from
ligand ions surrounding the actinide ion, the six-fold degeneracy
of $j$=5/2 is lifted, as shown in Fig.~2.
Note that in the $j$-$j$ coupling scheme, it is enough to define
the one-electron potential, which is deduced from the CEF level
scheme of corresponding $f^1$-electron system.
First we consider the one $f$-electron state in the AuCu$_3$-type
cubic crystal structure, since this is the mother structure of
HoCoGa$_5$-type tetragonal compound.
The effects of tetragonality will be discussed later.
For the case of cubic symmetry, we identify two eigen energies as
$-240B_4^0$ for the $\Gamma_7$ doublet and
$120B_4^0$ for the $\Gamma_8$ quartet,
where $B_4^0$ is a cubic CEF parameter.
For the following discussion, it is quite convenient to introduce
``pseudospin'' to distinguish the degenerate states of
the Kramers doublet and ``orbital'' to label the different
Kramers doublets in $\Gamma_8$ quartet.
In Fig.~3(a), we show the shape of the wavefuction of two
$\Gamma_8$ orbitals, $\Gamma_8^a$ and $\Gamma_8^b$.

\begin{figure}[t]
\begin{center}
\includegraphics[width=0.9\textwidth]{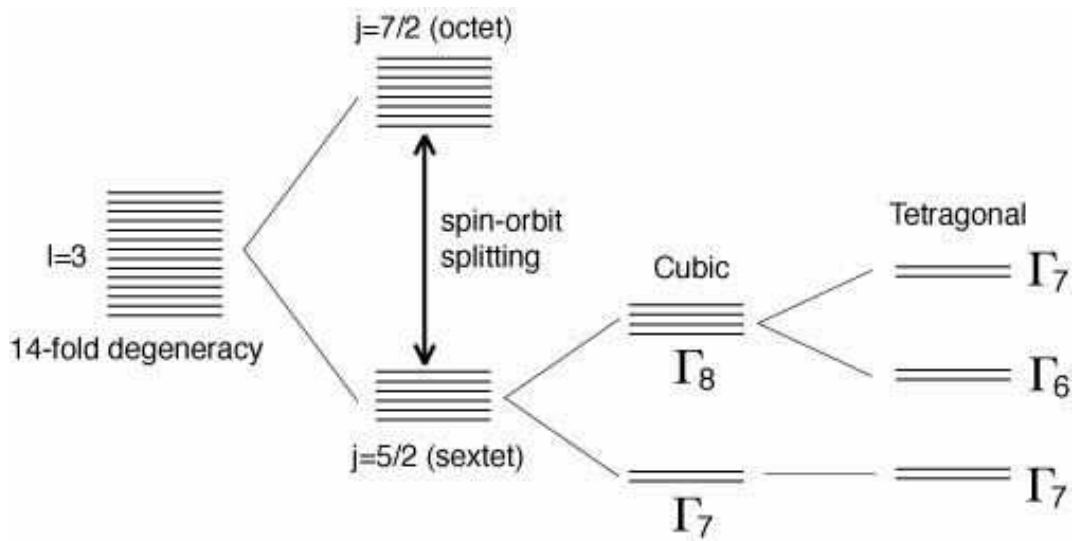}
\caption{%
Schematic view for local $f$-electron state in the $j$-$j$ coupling
scheme. The $j$=5/2 sextet is split into several Kramers doublets
due to cubic or tetragonal crystalline electric field effect.}
\end{center}
\end{figure}

In order to proceed with the discussion, we need to know
which is lower, $\Gamma_7$ or $\Gamma_8$,
in the one $f$-electron picture.
For some crystal structures, it is possible to determine
the level scheme from intuitive discussions on
$f$-electron wavefunctions and the positions of ligand ions.
However, this is not the case for the AuCu$_3$-type crystal structure.
For this case we invoke certain experimental results
on CeIn$_3$, a typical AuCu$_3$-type Ce-based compound,
where $\Gamma_7$ and $\Gamma_8$ have been reported as ground and excited
states, respectively, with an energy difference of 12meV \cite{Knafo}.
Thus, we take $\Gamma_7$ to be lower for the present considerations,
as shown in Fig.~3(b).

In the $j$-$j$ coupling scheme for UGa$_3$ and NpGa$_3$,
it is necessary to accommodate three or four electrons
in the one-electron energy states $\Gamma_7$ and $\Gamma_8$.
We immediately notice that there are two possibilities, i.e.,
low- and high-spin states, depending on the Hund's rule interaction
and the splitting between the $\Gamma_7$ and $\Gamma_8$ levels.
As discussed in Ref.~\cite{Hotta-jj},
the effective Hund's rule interaction becomes small
in the $j$-$j$ coupling scheme and thus,
the low-spin state should be realized,
as shown in Figs.~3(c) and (d).
We emphasize that the low-spin state is consistent with
the $LS$ coupling scheme.
In the electron configuration shown in Figs.~3(c) and (d),
the $\Gamma_7$ level is fully occupied to form a singlet.
If this $\Gamma_7$ level is located well below the $\Gamma_8$,
the occupying electrons will not contribute to the magnetic properties.
Thus, we ignore the $\Gamma_7$ electrons in the following discussion.
As for details to validate the suppression of $\Gamma_7$ level,
readers should refer Ref.~\cite{Hotta-U115}.
We also mention another theoretical effort based on the $j$-$j$ coupling
scheme, in which the dual nature, itinerant and localized, of $5f$ electron
has been emphasized \cite{Fulde1,Fulde2,Fulde3}.

\begin{figure}[t]
\begin{center}
\includegraphics[width=0.9\textwidth]{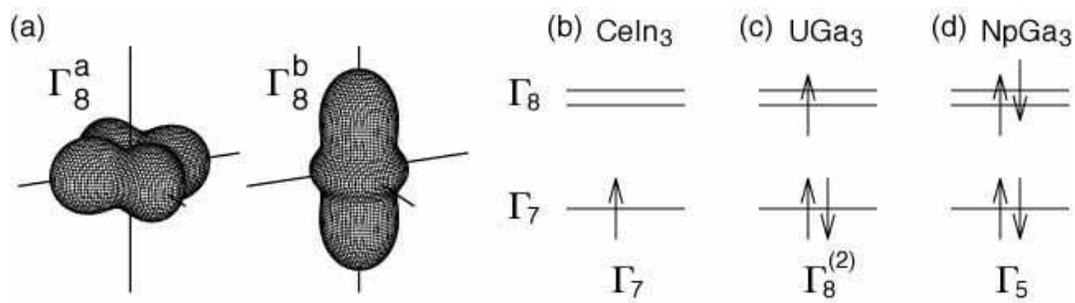}
\caption{%
(a) Two orbitals, $a$ and $b$, in the $\Gamma_8$ model.
Level scheme for (b) CeIn$_3$,
(c) UGa$_3$, and (d) NpGa$_3$ based on the $j$-$j$ coupling scheme.
Here we assume trivalent actinide ions as
U$^{3+}$ ($5f^3$) and Np$^{3+}$ ($5f^4$).
It should be noted that up and down arrows denote pseudospins to
distinguish the states in the Kramers doublet.
Note also that for NpGa$_3$, a couple of electrons in $\Gamma_8$
orbitals form a local triplet, leading to $\Gamma_5$.}
\end{center}
\end{figure}

As shown in Fig.~1(a), since 115 compounds have the tetragonal
structure, we need to include the effect of tetragonality.
In the tetragonal structure, the quartet $\Gamma_8$ is further split
into two doublets, $\Gamma_6$ and $\Gamma_7$, as shown in Fig.~2.
Namely, there appear three Kramers doublets with one $\Gamma_6$
and two $\Gamma_7$.
Note that two $\Gamma_7$ states are obtained from the mixture of
$\Gamma_7$ and $\Gamma_8^a$ in the cubic symmetry.
However, for simplicity, we ignore the change of wavefunctions
from cubic to tetragonal systems, since we believe that the existence
of orbital degree of freedom is a key issue
to understand the complex magnetic structure of actinide compound.
For the purpose to improve the present theory from the quantitative
viewpoint, it may be important to include also
the deformation of local $f$-electron wavefunction,
but we postpone such a study in the future.
In this paper, we consider the splitting energy
between $\Gamma_8$ orbitals, in order to take into account the
tetragonality in the 115 compounds.

\subsection{$\Gamma_8$ model}

After tedious calculations, we obtain the Hamiltonian including
$\Gamma_8$ orbitals in the form of \cite{Hotta-jj,Hotta-U115}
\begin{equation}
   H=H_{\rm kin}+H_{\rm CEF}+H_{\rm C},
\end{equation}
where $H_{\rm kin}$ denotes the kinetic term of $5f$ electrons,
$H_{\rm CEF}$ is the CEF potential term,
and $H_{\rm C}$ indicates the Coulomb interaction term among
$\Gamma_8$ electrons.

Concerning the kinetic term, one traditional way is to consider
the hybridization process between $f$ and conduction electrons.
For 115 systems, from the band-structure calculations \cite{Maehira},
$p$ electrons of Ga ion play an important role to form conduction band.
However, based upon a picture that $5f$ electrons are itinerant,
we can consider another way to construct the kinetic term
for $f$ electrons within a tight-binding approximation.
In actual compounds, there should occur several processes
through other ligand ions, in addition to the direct process
of $f$ electrons.
However, when we include correctly the local symmetry of
relevant $f$ orbitals, it is believed that we can grasp
qualitative point of kinetic term of $f$ electrons,
even within the simple tight-binding approximation.
This point will be discussed elsewhere in future.
Here we do not consider the hybridization process.
Then, the kinetic term is given in the tight-binding approximation
for $f$ electrons as
\begin{equation}
   H_{\rm kin} =
   \sum_{{\bf i,a},\sigma,\tau,\tau'} t^{\bf a}_{\tau\tau'}
   f^{\dag}_{{\bf i}\tau\sigma} f_{{\bf i+a}\tau'\sigma},
\end{equation}
where $f_{{\bf i}\tau\sigma}$ is the annihilation operator
for an $f$ electron with pseudospin $\sigma$ in the $\tau$-orbital
at site ${\bf i}$ and
$t^{\bf a}_{\tau\tau'}$ is the $f$-electron hopping matrix element
between $\tau$- and $\tau'$-orbitals along the ${\bf a}$ direction.
Indices $a$ and $b$ denote the $\Gamma_8^{a}$ and
$\Gamma_8^{b}$ orbitals, respectively.
In the $xy$ plane and along the $z$-axis,
$t^{\bf a}_{\tau\tau'}$ is given by
\begin{equation}
  \label{Eq:tx}
  t_{\tau\tau'}^{\bf x} = t
  \left(
   \begin{array}{cc}
     3/4 & -\sqrt{3}/4 \\
     -\sqrt{3}/4 & 1/4 \\
   \end{array}
  \right),
\end{equation}
for the ${\bf x}$-direction,
\begin{equation}
  \label{Eq:ty}
  t_{\tau\tau'}^{\bf y} = t
  \left(
   \begin{array}{cc}
     3/4 & \sqrt{3}/4 \\
     \sqrt{3}/4 & 1/4 \\
   \end{array}
  \right),
\end{equation}
for the ${\bf y}$ direction, and
\begin{equation}
  \label{Eq:tz}
  t_{\tau\tau'}^{\bf z} = 
  \left(
    \begin{array}{cc}
     0 & 0 \\
     0 & t_{bb}^{\bf z} \\
    \end{array}
  \right),
\end{equation}
for the ${\bf z}$ direction.
Note that $t$=(3/7)$(ff\sigma)$, where $(ff\sigma)$ is a Slater-Koster
overlap integral between $f$ orbitals through $\sigma$ bond.
In the following, $t$ is taken as an energy unity.
Remark that we introduce another hopping amplitude
along the $z$-axis.
In AnTGa$_5$ (An=U and Np), AnGa$_2$ layer is sandwiched by
two AnGa$_3$ sheets,
indicating that the hopping of $f$ electron along the $z$-axis
should be reduced from that in AnGa$_3$.
However, it is difficult to estimate the reduction quantitatively,
since it is necessary to include correctly the hybridization
with $d$ electrons in transition metal ions and $p$ electrons
in Ga ions.
Thus, in this paper, we consider the effective reduction
by simply treating $t_{bb}^{\bf z}$ as a parameter.
For a practical purpose, we introduce a non-dimensional
parameter $t_z$=$t_{bb}^{\bf z}/t$.

We point out that the hopping amplitudes among $\Gamma_8$ orbitals
are just the same as those for the $e_{\rm g}$ orbitals of
$3d$ electrons \cite{Dagotto,Hotta-CE}.
Readers can intuitively understand this point from the
shapes of $\Gamma_8$ orbitals shown in Fig.~3(a),
which are similar to $e_{\rm g}$ orbitals.
Mathematically, this is quite natural, if we recall the fact that
$\Gamma_8$ is isomorphic to $\Gamma_3 \times \Gamma_6$,
where $\Gamma_3$ indicates $E$ representation for the orbital part
and $\Gamma_6$ denotes the spin part.
Note, however, that the agreement between $f$- and $d$-electron
hopping amplitudes is due to the choice of special axis directions.
If we take the (1,1,0) direction in an fcc lattice,
$f$-electron hopping amplitude becomes complex,
leading to the stabilization of octupole ordering
in NpO$_2$ \cite{Kubo}.

The CEF term is given by
\begin{equation}
   H_{\rm CEF}=
   -\Delta \sum_{\bf i}(\rho_{{\bf i}a}-\rho_{{\bf i}b})/2,
\end{equation}
where
$\rho_{{\bf i}\tau\sigma}$=
$f_{{\bf i}\tau\sigma}^{\dag} f_{{\bf i}\tau\sigma}$,
$\rho_{{\bf i}\tau}$=$\sum_{\sigma}\rho_{{\bf i}\sigma\tau}$,
and $\Delta$ is the level splitting, which expresses
the effect of tetragonal CEF, as discussed above.
We note that in actuality, $\Delta$ should be related to
the value of $t_z$, since both quantities depend on
the lattice constant along the $z$ axis.
However, the relation between $t_z$ and $\Delta$
is out of the scope of this paper and thus,
here we simply treat them as independent parameters.

The Coulomb interaction term is expressed by
\begin{eqnarray}
 H_{\rm C} &=&
 U \sum_{{\bf i},\tau}\rho_{{\bf i}\tau\uparrow}
 \rho_{{\bf i}\tau\downarrow}
 + U' \sum_{\bf i} \rho_{{\bf i}a} \rho_{{\bf i}b} \nonumber \\
 &+& J \sum_{{\bf i},\sigma,\sigma'}
 f_{{\bf i}a\sigma}^{\dag} f_{{\bf i}b\sigma'}^{\dag}
 f_{{\bf i}a\sigma'} f_{{\bf i}b\sigma}
 + J' \sum_{{\bf i},\tau \ne \tau'}
 f_{{\bf i}\tau\uparrow}^{\dag} f_{{\bf i}\tau\downarrow}^{\dag}
 f_{{\bf i}\tau'\downarrow} f_{{\bf i}\tau'\uparrow},
\end{eqnarray}
where the Coulomb interactions $U$, $U'$, $J$, and $J'$ denote
intra-orbital, inter-orbital, exchange, and pair-hopping interactions
among $\Gamma_8$ electrons, respectively,
expressed by using the Racah parameters \cite{Hotta-acta}.
Note that the relation $U$=$U'$+$J$+$J'$ holds, ensuring rotational
invariance in pseudo-orbital space for the interaction part.
For $d$-electron systems, one also has the relation $J$=$J'$.
When the electronic wavefunction is real, this relation is
easily demonstrated from the definition of the Coulomb integral.
However, in the $j$-$j$ coupling scheme the wavefunction is
complex, and $J$ is not equal to $J'$ in general.
For simplicity, we shall assume here that $J$=$J'$,
noting that essential results are not affected.
Since double occupancy of the same orbital is suppressed
owing to the large value of $U$, pair-hopping processes
are irrelevant in the present case.

%
%
\section{Results}

Now let us show our calculated results based on the two-orbital model.
First we discuss the electronic properties of $H_{\rm kin}$
in the non-interacting limit by focusing on the Fermi-surface
structure, in order to gain an insight into
the interpretation of dHvA experiments on Np-115 compounds.
Next the magnetic properties are discussed.
We attempt to understand the appearance of three kinds of magnetic
phases based on the orbital-based scenario
similar to that of manganites.

\begin{figure}[t]
\begin{center}
\includegraphics[width=0.9\textwidth]{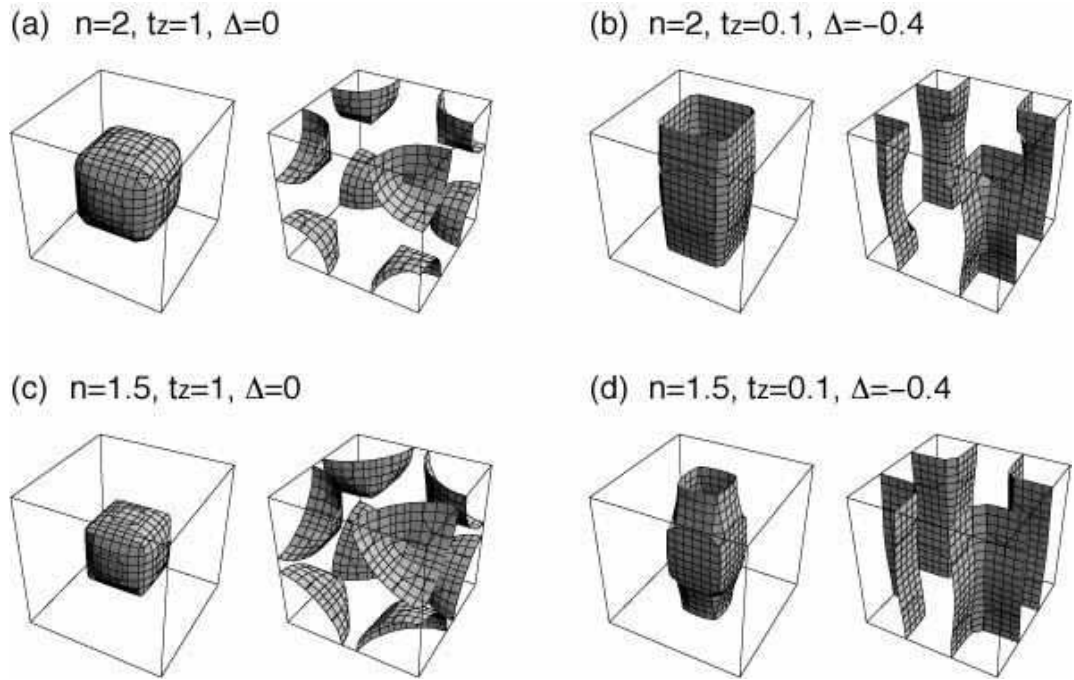}
\caption{%
Fermi-surface sheets of the $\Gamma_8$ tight-binding model for
(a) $n$=2, $t_z$=1, $\Delta$=0,
(b) $n$=2, $t_z$=0.1, $\Delta$=$-0.4$,
(c) $n$=1.5, $t_z$=1, $\Delta$=0, and
(d) $n$=1.5, $t_z$=0.1, $\Delta$=$-0.4$.
The bounding box indicates the first Brillouin zone
for a simple cubic lattice.
The $\Gamma$ point is located at the center of the box,
while the apices denote R points.}
\end{center}
\end{figure}

\subsection{Fermi-Surface Structure}

Let us consider the structure of Fermi-surface sheets of
the $\Gamma_8$ tight-binding model $H_{\rm kin}$.
In the following, we define $n$ as the number of $f$ electrons
included in the $\Gamma_8$ state.
Namely, the local $f$-electron number per actinide ion is
$n$+2, by adding $\Gamma_7$ electrons.

As shown in Fig.~3(d), in our picture, there are two $\Gamma_8$
electrons in trivalent neptunium ion.
Then, it is quite natural to consider first the case of $n$=2.
The results for ($t_z$, $\Delta$)=(1,0) and (0.1,$-0.4$) are
shown in Figs.~4(a) and (b), respectively.
For the cubic case with $t_z$=1 and $\Delta$=0,
there exist two kinds of cube-like Fermi surfaces.
One is centered at $\Gamma$ point, while the center of another
cube is R point.
When we change $t_z$ and $\Delta$ to consider the tetragonality,
cubes gradually change to cylinders.
As shown in Fig.~4(b), we can clearly observe two
kinds of cylindrical Fermi surfaces.
Note that the Fermi-surface structure does not sensitively depend on
the value of $\Delta$, as long as the absolute value is
not so large as $|\Delta|$$<$0.5.
It is important to have quasi orbital degeneracy,
to reproduce a couple of cylindrical Fermi surfaces.

It is interesting to recall the dHvA experiment on NpCoGa$_5$,
which has revealed two kinds of cylindrical Fermi surface
\cite{Aoki-Co2}.
In the relativistic band-structure calculations
for the paramagnetic phase,
it has been difficult to understand the appearance of a couple of
cylindrical Fermi surfaces \cite{Maehira}.
Note that the folding of the magnetic Brillouin zone cannot
explain the discrepancy between the experimental results and
the band-calculation ones for NpCoGa$_5$.
A direct way to understand the dHvA experimental results is
to improve the band-structure calculation method, but
as shown in Fig.~4(b), even in a simple tight-binding model
for itinerant $5f$ electrons,
we obtain two kinds of cylindrical Fermi surface
due to the effect of $\Gamma_8$ orbital degree of freedom.
We do not claim that the dHvA results can be understood only
by the present simple calculations, but a hint to understand
the electronic structure of Np-115 is believed to be hidden in
the construction of our model
with $\Gamma_8$ orbital degree of freedom.

We point out that in actual Np-115 compounds,
the number of $n$ is not definitely determined.
Especially, due to the change of transition metal ion,
the number of $n$ seems to be changed effectively.
In fact, in the relation between superconducting transition
temperature $T_{\rm c}$ and the ratio of lattice constants $c/a$,
the curve of $T_{\rm c}$ vs. $c/a$ for solid solution (U,Np,Pu)CoGa$_5$
is similar to that for another solid solution Pu(Fe,Co,Ni)Ga$_5$
\cite{Colineau2,Boulet}.
Namely, in actinide 115 materials,
the effect of the change of $f$-electron number
due to the substitution of actinide ion
seems to be similar to that of
the substitution of transition metal ion.
Thus, when we consider the change of the magnetic structure among
NpTGa$_5$ (T=Fe, Co, and Ni) based on the effective two-orbital model,
there is no strong reason to fix firmly $n$=2.

In Figs.~4(c) and (d), we show the Fermi-surface sheets for $n$=1.5,
in which the number of $f$ electrons is slightly decreased.
We can observe that the Fermi-surface structure is not changed so much,
although the cylinder centered at the $\Gamma$ point becomes slender.
Thus, as long as we are based on the simple tight-binding model,
it is concluded that the structure of the Fermi-surface sheets
is not changed sensitively by the number of $n$ around at $n$=2.
This result suggests that it is possible to change the value of $n$
to consider the magnteic structure of Np-115 compounds.

\subsection{Magnetic structure}

Now we consider the magnetic properties of the $\Gamma_8$ model.
Among several methods to analyze the microscopic model,
in this paper we resort to an exact diagonalization technique
on a 2$\times$2$\times$2 lattice.
Although there is a demerit that it is difficult to enlarge the
system size, we take a clear advantage that it is possible to
deduce the magnetic structure
by including the effect of electron correlation.
In order to discuss the ground-state properties,
it is useful to measure the spin and orbital correlations,
which are, respectively, defined by
\begin{equation}
  S(\bm{q}) = (1/N)\sum_{{\bf i,j}}
  \langle \sigma^{z}_{\bf i} \sigma^{z}_{\bf j} \rangle
  e^{i\bm{q}\cdot({\bf i}-{\bf j})},
\end{equation}
with $\sigma^{z}_{\bf i}$=%
$\sum_{\tau}(n_{{\bf i}\tau\uparrow}$$-$$n_{{\bf i}\tau\downarrow})/2$,
and
\begin{equation}
  T(\bm{q}) = (1/N)\sum_{{\bf i,j}}
  \langle {\bf \tau}^{z}_{\bf i}{\bf \tau}^{z}_{\bf j} \rangle
  e^{i\bm{q}\cdot({\bf i}-{\bf j})},
\end{equation}
with $\tau^{z}_{\bf i}$=%
$\sum_{\sigma}(n_{{\bf i}a\sigma}$$-$$n_{{\bf i}b\sigma})/2$.
Here $N$ is the number of sites.

\begin{figure}[t]
\begin{center}
\includegraphics[width=0.9\textwidth]{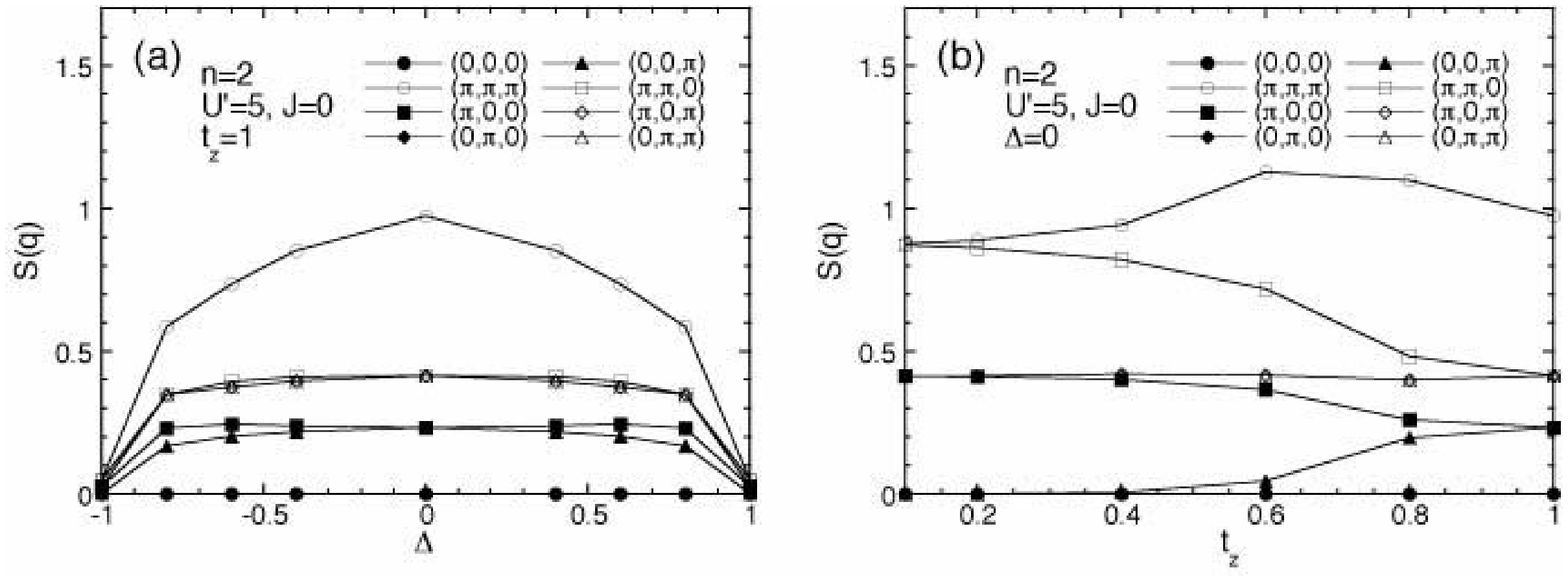}
\caption{%
Spin correlation for $n$=2, $U'$=5, and $J$=0
as a function of (a) $\Delta$ for $t_z$=1
and of (b) $t_z$ for $\Delta$=0.}
\end{center}
\end{figure}

First let us consider the case of $n$=2,
which is corresponding to the trivalent Np ion.
In Fig.~5(a), we show the spin correlation as a function
of $\Delta$ for $U'$=5, $J$=0, and $t_z$=1.
At $t_z$=1 and $\Delta$=0 (cubic system),
local triplet composed of a couple of $f$ electrons is formed
at each site and the G-type AF structure is stabilized
due to the so-called superexchange interaction.
As easily observed, even when $\Delta$ is introduced
as the tetragonal CEF effect,
the G-AF structure remains robust for $|\Delta|<1$.
When $|\Delta|$ is larger than unity, two electrons simultaneously
occupy the lower orbital, leading to the non-magnetic state
composed of local $\Gamma_1$, irrelevant to the present study
to consider the magnetic phase.
In Fig.~5(b), the spin correlation is shown
as a function of $t_z$ for $\Delta$=0.
Again, the G-type AF structure is stabilized,
but we find that the spin correlation of ${\bm q}$=$(\pi,\pi,0)$
comes to be equivalent to that of ${\bm q}$=$(\pi,\pi,\pi)$
with the decrease of $t_z$,
since the AF structure is stabilized in each $xy$ plane
due to superexchange interaction
and the planes are decoupled for small $t_z$.

At the first glance, it seems difficult to understand the variety of
magnetic phases observed in NpTGa$_5$, when we consider only
the trivalent Np ion.
However, there is no {\it a priori} reason to fix the valence as
Np$^{3+}$, as mentioned in the previous subsection.
In NpTGa$_5$, $d$-electron band originating from transition metal
ions may significantly affect the valence of Np ion.
In addition, we also stress that the actual compounds exhibit
AF metallic behavior. In the band-structure calculation,
the average number of $f$ electrons at Np ion is easily
decreased from four.
Thus, we treat the local $f$-electron number as a parameter.

We may consider another reason to decrease effectively the number of
$f$ electron from $n$=2 in NpGa$_3$.
In the present two-orbital model, the G-AF structure is robust,
which is natural from the theoretical viewpoint within the model.
However, in the experimental result on NpGa$_3$,
the low-temperature ground state is ferromagnetic, although
the AF phase has been observed around at $T$$\sim$60K \cite{NpGa3}.
In order to understand the occurrence of the FM phase in the
two-orbital model, it is necessary to inject some amount of
``hole'' in the AF phase, since the double-exchange mechanism works
to maximize the kinetic motion of electrons, as will be discussed
later in the main text.
It is difficult to determine the amount of doped holes
to obtain the FM phase, but at least qualitatively,
the effective decrease of $n$ seems to be physically meaningful
in NpGa$_3$ as well as in NpTGa$_5$.

Now we consider the case of $n$=1.5.
As discussed in the previous subsection, the Fermi-surface
structure does not change so much from that of the case of $n$=2,
when $|\Delta|$ is not so large.
In Fig.~6(a), we show the ground-state phase diagram
in the $(U',J)$ plane at $n$=1.5 for the cubic case
with $t_z$=1 and $\Delta$=0.
At $J$=0, a G-type AF structure is stabilized due to
superexchange interaction in the same way as the case of $n$=2.
However, the G-AF structure is immediately changed to a C-AF structure
only by a small value of the Hund's rule interaction.
With increasing $J$, the magnetic phase changes
in the order of G-AF, C-AF, A-AF, and FM phases,
except for the C-AF phase in the large $J$ region.
The spin structure of each phase is shown in Fig.~6(b).
This result is quite natural, since we are now considering
the magnetic structure based on the two-orbital model,
in which the FM tendency is due to the optimization
of kinetic motion of electrons.

\begin{figure}[t]
\begin{center}
\includegraphics[width=0.8\textwidth]{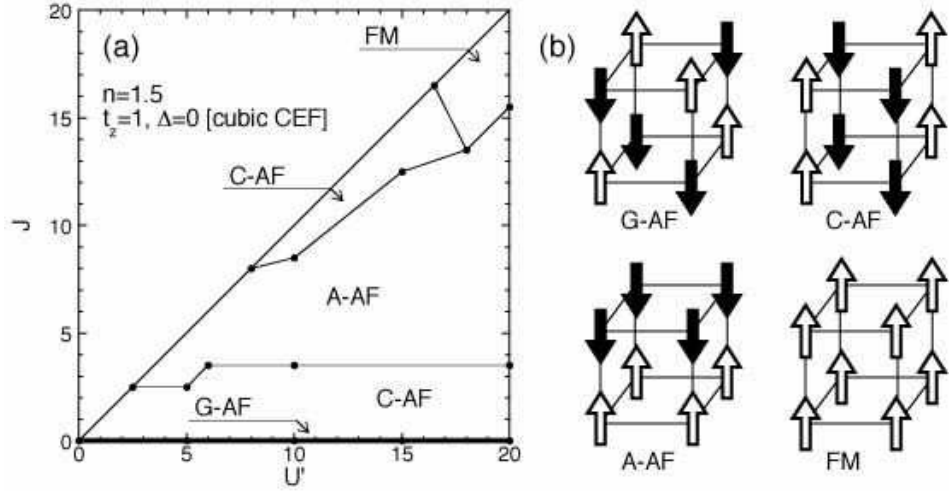}
\caption{%
(a) Ground-state phase diagram in the $(U',J)$ plane
for $n$=1.5, $t_z$=1, and $\Delta$=0.
(b) Spin structures in the G-type AF, C-type AF, A-type AF,
and FM phases.}
\end{center}
\end{figure}

In order to understand the underlying mechanism of the appearance
of various magnetic structures in addition to the G-AF structure
for $n$=1.5, we consider the case of $J$=0.
In Fig.~7(a), we show typical results of the spin correlation
as a function of $\Delta$ for $U'$=5, $J$=0, and $t_z$=1.
At $\Delta$=0, the dominant component of $S({\bm q})$ appears
at ${\bm q}$=$(\pi,\pi,\pi)$, indicating the G-type AF structure.
On the other hand, in the region of positive small $\Delta$,
the spin correlation of ${\bm q}$=$(\pi,\pi,0)$ turns to be dominant,
indicating the C-type AF structure.
This phase is defined as C-AF(I).
Concerning the orbital correlation,
we find no remarkable structure in the C-AF(I) phase,
as shown in Fig.~7(b).
Note that in the G-AF phase with positive $\Delta$,
the orbital correlation of ${\bm q}$=$(0,0,0)$ has a significant value
in comparison with other regions,
since electrons tend to doubly occupy lower $\Gamma_8^a$ orbitals.
This G-AF phase is called G-AF(I).
On the other hand, in the G-AF phase with negative $\Delta$,
there is no indication of the development of
the orbital correlation of ${\bm q}$=$(0,0,0)$,
since the lower $\Gamma_8^b$ orbitals are singly occupied.
We label this G-AF phase as G-AF(II) to distinguish it from G-AF(I).
It is interesting to see different types of the G-AF phases
due to the change of the orbital state.

Let us now discuss the reason of the appearance of the C-AF(I) phase
in the region of positive small $\Delta$.
For $\Delta$$>$0, the energy level of $\Gamma_8^a$ is lower
than that of $\Gamma_8^b$ by definition.
Then, the $\Gamma_8^a$ orbital is first occupied
by one electron at each site.
The rest of electrons are accommodated in $\Gamma_8^b$ orbitals
to avoid the effect of intra-orbital Coulomb interaction
when $\Delta$ is not large so much.
The electrons in $\Gamma_8^a$ orbitals hop to the $\Gamma_8^a$ orbital
at the nearest-neighbor site, since the $\Gamma_8^b$ has higher energy.
Since the electrons in $\Gamma_8^a$ orbitals move only in the $xy$ plane,
the AF structure is stabilized in each $xy$ plane independently
due to superexchange interaction, as schematically shown in Fig.~7(c).
On the other hand,
the electrons in $\Gamma_8^b$ orbitals can move itinerantly
within the network composed of $\Gamma_8^b$ orbitals
due to the existence of holes in $\Gamma_8^b$ orbitals.
Since the hopping amplitude along the $z$ axis is dominant
in the region near $t_z$=1,
the motion of $\Gamma_8^b$ electrons causes an FM spin arrangement
of $\Gamma_8^a$ electrons along the $z$ axis to gain kinetic energy.
Namely, even though localized $t_{\rm 2g}$ spins do not exist
in the present case, the so-called double-exchange mechanism
is always active in the two-orbital model.
Note that the C-AF structure disappears for large $\Delta$,
since all the electrons tend to occupy $\Gamma_8^a$ orbitals
and the double-exchange mechanism is no longer active in that situation.

\begin{figure}[t]
\begin{center}
\includegraphics[width=0.9\textwidth]{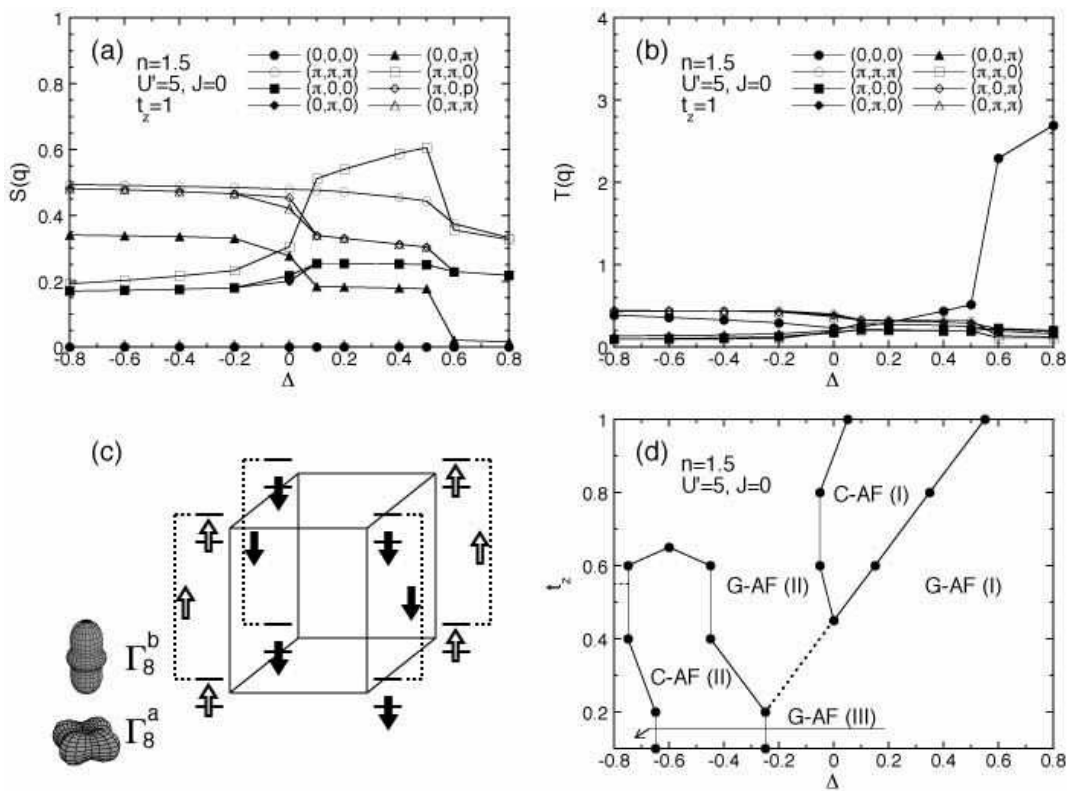}
\caption{%
(a) Spin and (b) orbital correlations
as a function of $\Delta$ for $n$=1.5, $U'$=5 $J$=0, and $t_z$=1.
(c) Schematic view of the electron configuration
in the C-AF phase in the region near $t_z$=1.
(d) Ground-state phase diagram of the magnetic structure
in the $(\Delta,t_z)$ plane.}
\end{center}
\end{figure}

After calculations of the spin and orbital correlations
for several parameter sets, we obtain the ground-state phase diagram
in the $(\Delta,t_z)$ plane, as shown in Fig.~7(d).
It is found that the C-AF(I) phase discussed above extends to
a wide region of $t_z$$<$1 in the phase diagram.
Note that for small values of $t_z$,
another C-AF phase, which we call C-AF(II), appears
in the region of negative small $\Delta$,
in which the spin correlations of
$(\pi,0,\pi)$ and $(0,\pi,\pi)$ are dominant.
For small $t_z$ and large negative $\Delta$, there appears
yet another G-AF phase, called G-AF(III),
due to the same reason as that for the occurrence of G-AF(I)
phase for positive $\Delta$, but the occupied orbital is changed
from $\Gamma_8^a$ to $\Gamma_8^b$.
In any case, for $J$=0, we can observe several kinds of C- and G-AF
phases, but A-AF phase does not occur.

\begin{figure}[t]
\begin{center}
\includegraphics[width=0.9\textwidth]{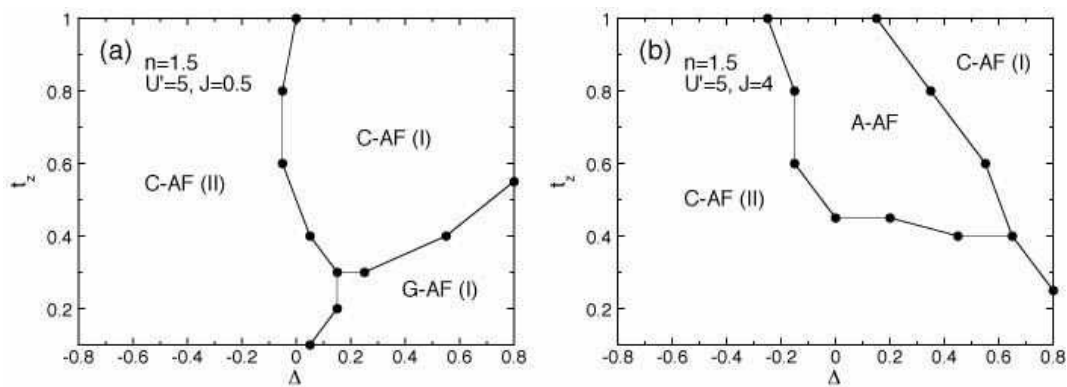}
\caption{%
Ground-state phase diagram of the magnetic structure
in the $(\Delta,t_z)$ plane for $n$=1.5, $U'$=5,
(a) $J$=0.5 and (b) $J$=4.}
\end{center}
\end{figure}

Although we increase the value of $J$ as $J$=0.5, no new phases
appear in the phase diagram for $n$=1.5, as shown in Fig.~8(a).
There are three phases, but they are two C-AF and one G-AF states.
As labelled explicitly in the phase diagrams, C-AF(I), C-AF(II),
and G-AF(I) are the same as those in the phase diagram of Fig.~7(d).
Due to the effect of $J$, G-AF(II) and G-AF(III) disappear,
since the number of FM bonds should be increased to gain
the kinetic energy.
As shown in Fig.~8(b), when we further increase the value of $J$
as $J$=4, the G-AF phase completely disappears and instead,
we observe the A-AF phase sandwiched by two C-AF phases.
As described above, due to the double-exchange mechanism
in the two-orbital model, the A-AF phase is considered to appear,
when $J$ is increased.

%
%
\section{Discussion and Summary}

In this paper, we have proposed an effective model with
active orbital degree of freedom to understand
the magnetic structure of neptunium 115 compounds from the
microscopic viewpoint.
By analyzing the model with the use of
the exact diagonalization technique,
we have obtained the ground-state phase diagrams
including three kinds of AF phases 
corresponding to NpTGa$_5$.

In the experiments for NpTGa$_5$,
C-, A-, and G-AF magnetic phases have been found in
NpFeGa$_5$, NpCoGa$_5$, and NpNiGa$_5$.
Here we have a question:
What is a key parameter to understand
the change of the magnetic structure?
In the case of UTGa$_5$, it has been claimed
that the level splitting $\Delta$ is important
to explain the difference in magnetic structure as
well as the magnetic anisotropy for a fixed value of $n$=1
\cite{Hotta-U115}.
Roughly speaking, $\Delta$ is positive for T=Fe,
small positive for T=Co, and negative for T=Ni.
Among UTGa$_5$ with T=Ni, Pd, and Pt, when we assume that
the absolute value of $\Delta$ is increased in the order
of Ni, Pd, and Pt, it is possible to understand qualitatively
the change in the magnetic anisotropy,
in addition to the change in the magnetic structure of
G-AF for T=Ni and A-AF for T=Pd and Pt.
It has been found that the value of $t_z$ is not so crucial
to explain qualitatively the magnetic properties of
U-115 based on the two-orbital model for $n$=1.

For $n$=2, as emphasized in the previous section,
we always obtain the G-AF phase.
However, for $n$=1.5, we have observed three kinds of AF
magnetic structure in the phase diagrams.
Let us summarize the change in the magnetic structure
for a fixed value of $t_z$=0.8.
Note that this value is larger than $t_z$=0.1, which we have
considered to reproduce two kinds of cylindrical Fermi-surface
sheets of Np-115.
However, in the small-sized cluster calculations,
it is difficult to compare directly with the values in
the thermodynamic limit.
Thus, we do not discuss further the quantitative point on
the values of $t_z$ here.
As shown in Fig.~7(d), for $J$=0 and $t_z$=0.8,
we see the change in the magnetic structure
as G-AF ($\Delta$$<$0), C-AF(0$<$$\Delta$$<$0.4),
and G-AF ($\Delta$$>$0.4).
For $J$=0.5 and $t_z$=0.8, as shown in Fig.~8(a),
the C-AF phases are always observed, but they have
different orbital structures.
Finally, for $J$=4 and $t_z$=0.8, we observe
C-AF ($\Delta$$<$-0.15), A-AF(-0.15$<$$\Delta$$<$0.3),
and C-AF ($\Delta$$>$0.3), as shown in Fig.~8(b).

In order to understand the appearance of three types of
the AF phases, we may consider an explanation due to
the combination of the changes in $\Delta$ and $n$.
For instance, by assuming that $J$=4 for NpTGa$_5$
and the change in $\Delta$ for NpTGa$_5$ is just
the same as that for UTGa$_5$,
we consider that $n$$\approx$2 with $\Delta$$<$0 for T=Ni,
$n$$\approx$1.5 with $\Delta$$\approx$0 for T=Co,
and $n$$\approx$1.5 with $\Delta$$>0$ for T=Fe.
Then, it seems to be possible to relate our theoretical AF phases
with the experimental observations in NpTGa$_5$.
However, it is difficult to claim that the above parameter assignment
for three Np-115 materials is the best explanation for
the magnetic structure of Np-115,
since in actual compounds, there are other important ingredients
which have not been included in the present model.
For instance, we have never discussed the direction of the
magnetic moment of Np ion.
Especially, the canted AF structure cannot be considered at all
for the G-AF phase of NpNiGa$_5$.
Thus, we need to recognize some distance between the actual magnetic
states and the theoretically obtained phases.
Our theory should be improved by taking into account
other ingredients of 115 structure.

In summary, we have analyzed the orbital degenerate model
appropriate for NpTGa$_5$ by using the exact diagonalization
technique.
Our phase diagram includes C-, A-, and G-AF phases.
We have proposed the manganite-like scenario to understand
the appearance of three kinds of AF spin structure in Np-115.
Namely, the double-exchange mechanism works also in some
actinide compounds based on the model with active orbital
degree of freedom.
We believe that the present model can grasp some important points
of the actinide compound by regarding it as charge-spin-orbital complex.

\section*{Acknowledgement}

The authors thank D. Aoki, Y. Haga, H. Harima, Y. Homma, F. Honda,
S. Ikeda, S. Kambe, K. Kaneko, K. Kubo, T. Maehira, T. D. Matsuda,
N. Metoki, A. Nakamura, Y. \=Onuki, H. Sakai, Y. Shiokawa,
T. Takimoto, Y. Tokunaga, K. Ueda, R. E. Walstedt, F. Wastin,
E. Yamamoto, and H. Yasuoka for discussions.
This work has been supported by Grant-in-Aids for Scientific
Research (No.~14740219)
of Japan Society for the Promotion of Science
and for Scientific Research in Priority Area ``Skutterudites''
(No.~16037217) of the Ministry of
Education, Culture, Sports, Science, and Technology of Japan.
The computation in this work has been done using the facilities
of the Supercomputer Center, Institute for Solid State Physics,
University of Tokyo.


\end{document}